\documentclass[12pt]{iopart}
\def\IGNORE#1{}
\usepackage{iopams}
\usepackage{epsfig}

\def\hb{\hfil\break}

\makeatletter

\DeclareRobustCommand{\binom}{\frbinom{}}
\def\frbinom#1#2#3{{#1{#2\atopwithdelims()#3}}}
\makeatother

\begin{document}

\title[The low autocorrelated 
binary string problem]
{Landscape statistics  of the  low autocorrelated 
binary string problem }

\author{Fernando F.\ Ferreira$^{a}$,
        Jos\'e F.\ Fontanari$^{a}$, and 
        Peter F.\ Stadler$^{b,c}$ \footnote[3]{To
           whom correspondence should be addressed.\\
           Email: {\tt studla@tbi.univie.ac.at},
           Phone: **43 1 4277 52737,
           Fax: **43 1 4277 52793}
}

\address{$^a$Instituto de F{\'\i}sica de S{\~a}o Carlos,
            Universidade de S{\~a}o Paulo,
            Caixa Postal 369, 13560-970 S\~ao Carlos SP, Brazil}

\address{$^b$Institut f{\"u}r Theoretische Chemie, Universit{\"a}t Wien
            W{\"a}hringerstra{\ss}e 17, A-1090 Wien, Austria}

\address{$^c$The Santa Fe Institute, 
            1399 Hyde Park Road, Santa Fe, NM 87501, USA}

\begin{abstract}

The statistical properties of the energy landscape of the low
autocorrelated binary string problem ({\tt LABSP}) are studied numerically
and compared with those of several classic disordered models. Using two
global measures of landscape structure which have been introduced in the
Simulated Annealing literature, namely, depth and difficulty, we find that
the landscape of {\tt LABSP}, except perhaps for a very large degeneracy of
the local minima energies, is qualitatively similar to some well-known
landscapes such as that of the mean-field 2-spin glass model.  Furthermore,
we consider a mean-field approximation to the pure model proposed by
Bouchaud and M\'ezard (1994, {\it J.\ Physique I France} {\bf 4} 1109) and
show both analytically and numerically that it describes extremely well the
statistical properties of {\tt LABSP}.

\end{abstract} 


\maketitle


\section{Introduction}
\label{sect:intro}

The Low Autocorrelated Binary String Problem ({\tt LABSP})
\cite{Golay:77,Bernasconi:87} consists of finding binary strings $x$ of
length $N$ over the alphabet $\{\pm1\}$ with low aperiodic off-peak
autocorrelation $R_k(x) = \sum_{i=1}^{N-k} x_i x_{i+k}$ for all lags
$k$. These strings have technical applications such as the synchronization
in digital communication systems and the modulation of radar pulses. 

The quality of a string
$x$ is measured by the fitness or energy function
\begin{equation}\label{eq:H}
{\mathcal{H}}(x) = \frac{1}{2N} \sum_{k=1}^{N-1} \left [ \sum_{i=1}^{N-k}
x_i x_{i+k} \right]^2 = \frac{1}{2N} \sum_{k=1}^{N-1} R_k(x)^2\,.
\end{equation}
In most of the literature on the {\tt LABSP} the {\it merit factor}
$F(x)= N^2/(4{\mathcal{H}}(x))$ is used (see e.g.\ \cite{Bernasconi:87}):
using ${\mathcal{H}}$ instead is more convenient for explicit computations.

Recently there has been much interest in frustrated models without explicit
disorder. The {\tt LABSP} and related bit-string problems have served as
model systems for this avenue of research
\cite{Marinari:94a,Marinari:94b,Migliorini:94,Bouchaud:94}. These
investigations have lead to a claim that {\tt LABSP} has a `golf-course' 
type landscape structure, which would explain the fact that it has been 
identified as a
particularly hard optimization problem for heuristic algorithms such as
Simulated Annealing (see \cite{Bernasconi:87,Militzer:98,Mertens:96} and 
the references
therein).

The landscape of {\tt LABSP} consists of a (dominant) 4-spin
Hamiltonian plus an asymptotically negligible quadratic component.  We note
that the generic $4$-spin landscape is Derrida's $4$-spin Hamiltonian
\cite{Derrida:80a} which is a linear combination of all $\binom{N}{4}$
distinct $4$-spin functions, while the {\tt LABSP} Hamiltonian, on the
other hand, only contains $\mathcal{O}(N^3)$ non-vanishing $4$-spin
contributions. The landscape of the {\tt LABSP} thus corresponds to a
dilute 4-spin ferromagnet. Numerical simulations in \cite{Oliveira:99b}
show that the {\tt LABSP} has by far more local optima than a generic
4-spin glass model, which corroborates the rather surprising finding that
disordered ferromagnets have more metastable states than their spin-glass
counterparts \cite{Cieplak:87}.

In this contribution we carry out a thorough investigation of the
statistical properties of the energy landscape of {\tt LABSP} aiming at to
determine whether it has any peculiar features that would lead to a
`golf-course' structure, with vanishingly small correlations between the
energies of neighboring states.  To do so we carry out a comparison with
four disordered models, namely, the random energy model ({\tt
REM})\cite{Derrida:80a}, the $\pm 1$ 4-spin glass model
\cite{Gardner:85,Oliveira:99a}, a mean-field approximation to
${\mathcal{H}}$ ({\tt MF}) \cite{Bouchaud:94}, which reproduces the results
of Golay's ergodicity assumption \cite{Golay:77}, and, finally, the $\pm 1$
2-spin glass model \cite{Sherrington:75}. The replica analyses indicate
that the first three models have a rather unusual spin-glass phase, where
the overlap between any pair of different equilibrium states vanishes,
while the last model has a normal spin-glass phase described by a
continuous order parameter function.

The rest of this paper is organized in the following way.
In section
\ref{sect:BM} we calculate analytically the average density of local minima
of the disordered mean-field approximation to ${\mathcal{H}}$ and show that 
it indeed  describes very well the statistics of metastable states 
of the pure model. Rather surprisingly, we find that the value of the 
energy density at which the 
density of local minima vanishes coincides with the bound predicted
by Golay \cite{Golay:77}, as well as with the ground-state energy predicted by the
first step of replica-symmetry breaking calculations of the mean-field model 
\cite{Bouchaud:94}. 
To properly compare the landscapes of the different
models mentioned above, in section \ref{sect:barriers} we consider two
global measures  of landscape structure which have been introduced
in the Simulated Annealing literature: depth and difficulty 
\cite{Hajek:88,Catoni:92,Kern:93,Ryan:95}. We show that
{\tt LABSP}, the mean-field approximation, and the binary $\pm 1$ 2 and 4-spin 
glasses exhibit approximately the same qualitative behavior in these parameters,
while the behavior pattern of
the random energy model departs significantly from those. 
Finally, in section \ref{sect:disc} we summarize our main results and
present some concluding remarks.

\section{Mean-field approximation}
\label{sect:BM}

Bouchaud and M\'ezard \cite{Bouchaud:94} and, independently, Marinari {\it
et al.}\ \cite{Marinari:94a} have proposed the following disordered model,
which is ``as close as possible'' to the pure model:
\begin{equation}\label{eq:H_BM}
{\mathcal{H}}_d = \frac{1}{2N} \sum_{k=1}^{N-1} \left[ \sum_{i=1}^{N}
\sum_{j \neq i}^N J_{ij}^{k} x_i x_j \right]^2
\end{equation}
Here the coupling strengths $J_{ij}^{k} \neq J_{ji}^{k}$ are statistically
independent random variables that can take on the value $1$ with
probability $\left( N - k \right)/N^2$ and zero otherwise. Hence the
average number of bonds in ${\mathcal{H}}$ and ${\mathcal{H}}_d$ is the
same, namely, $N-k$. Moreover, the  pure model is recovered with
the choice $J_{ij}^k = \delta_{i+k,j}$. Probably the most appealing feature
of this model is that its high-temperature (replica-symmetric) free-energy
is identical to that obtained by Bernasconi \cite{Bernasconi:87} using
Golay's ergodicity assumption \cite{Golay:77}, in which the squared
autocorrelations $R_k^2$ are treated as independent random variables. As
the constraints of the one-dimensional geometry are lost in the disordered
Hamiltonian ${\mathcal{H}}_d$, it can be viewed as the mean field version
of ${\mathcal{H}}$.

The thermodynamics of the disordered model (\ref{eq:H_BM}) is interesting
on its own since, similarly to the random energy model \cite{Derrida:80a},
it presents a first order transition at a certain temperature $T_g$, below
which the overlap between any pair of different equilibrium states
vanishes \cite{Bouchaud:94}. In contrast to the random energy model, 
however, the degrees of
freedom are not completely frozen for $T < T_g$, and the entropy vanishes
linearly with $T$ as the temperature decreases towards zero. To better
understand the low-temperature phase of the mean-field Hamiltonian
${\mathcal{H}}_d$, in the following we will calculate analytically the
expected number of metastable states $\langle\mathcal{N} \left ( \epsilon
\right) \rangle$ with a given energy density $\epsilon$.

The energy cost per site of 
flipping the spin $x_i$ is $\delta {\mathcal{H}}_d^i = - \Delta_i$ where
\begin{equation}
\label{delta}
\Delta_i = \sum_k v_i^k \left( \sum_j v_j^k - 2 v_i^k \right) 
\end{equation}
with
\begin{equation}
v_i^k =\frac{1}{\sqrt{N}} \sum_{j \neq i}
            \left( J_{ij}^k + J_{ji}^k \right) x_i x_j \,.
\end{equation}
We say that a state $x=\left(x_1,\ldots,x_N\right)$ is a strict 
local minimum if $\Delta_i < 0$ for all $i$; in the case that
the equality $\Delta_i = 0$ holds for some $i$,
we call $x$ a degenerate local minimum.
In the forthcoming analysis, the
choice of $\le$ instead of $<$, which is customary in optimization theory,
see e.g. \cite{Ryan:95}, does not make any difference. In
section~\ref{sect:barriers}, however, degeneracies will play a role.

The average number of
local minima with energy density $\epsilon$ can be written as
\begin{equation}\label{n1}
\langle {\mathcal{N}} \left( \epsilon \right) \rangle  =  
\left\langle \mbox{Tr}_x 
\, \delta \left[ \epsilon -  \frac{1}{N} {\mathcal{H}}_d \left(x\right) 
\right] \prod_i \Theta \left( -\Delta_i \right)  \right\rangle  
\end{equation}
where $\mbox{Tr}_x$ denotes the summation over the $2^N$ spin
configurations and $\langle\ldots\rangle$ stands for the average over the
couplings $J_{ij}^k$. Here $\Theta\left(x\right) = 1$ if $x > 0$ and $0$
otherwise, and $\delta\left(x\right)$ is the Dirac delta-function.

Using the integral representation of the delta-function we obtain
\begin{eqnarray}\label{n2}
\langle {\mathcal{N}} \left( \epsilon \right) \rangle &  = &
N \int \frac{d \hat{\epsilon}}{2 \pi}
  \mbox{e}^{{\mathbf i}N \hat{\epsilon} \epsilon}
\prod_i \int \frac{d \Delta_i d \hat{\Delta}_i}{2 \pi} \, 
\Theta \left( - \Delta_i \right)
  \mbox{e}^{ {\mathbf i} \hat{\Delta}_i \Delta_i}
\prod_{ik} \int \frac{dv_i^k d\hat{v}_i^k}{2 \pi}
  \mbox{e}^{{\mathbf i} v_i^k \hat{v}_i^k} 
\nonumber \\ 
&  &  \times 
\exp \left\{ 
- {\mathbf i} \frac{\hat{\epsilon}}{8} \sum_k \left(\sum_i v_i^k\right)^2  
- {\mathbf i} \sum_{ik} \hat{\Delta}_i v_i^k
              \left(\sum_j v_j^k - 2v_i^k \right)  \right\}
\nonumber \\
&  & \times \mbox{Tr}_x \left\langle
     \exp\left[-\frac{{\mathbf i}}{\sqrt{N}}
               \sum_{ik} \hat{v}_i^k \sum_{j \neq i}
                  \left( J_{ij}^k + J_{ji}^k \right)
               x_i x_j \right]
\right\rangle\, .
\end{eqnarray}
The average over the couplings can be easily carried out and, in
the thermodynamic limit $N \rightarrow \infty$, it yields
\begin{eqnarray}\label{average}
\ln \langle \ldots \rangle & = & - \sum_k  
\left( 1 - \frac{k}{N} \right) \left[ \frac{2{\mathbf i}}{\sqrt{N}} 
\left( \frac{1}{\sqrt{N}}
\sum_i x_i \right)
\left( \frac{1}{\sqrt{N}} \sum_i \hat{v}_i^k x_i \right) \right.
\nonumber \\
&  & + \left.
\frac{1}{N} \sum_i 
\left( \hat{v}_i^k \right) ^2 + \left( \frac{1}{N} \sum_i \hat{v}_i^k \right)^2
\right]  .
\end{eqnarray}
We note that this result could have been obtained by considering the
couplings $J_{ij}^k$ as Gaussian independent random variables with means
and variances equal to $\left(1-k/N\right)/N$. To get a physical but
nontrivial thermodynamic limit we must assume that the magnetization
$\sum_i x_i$ scales with $N^{1/2}$, which results then in the vanishing of
the term that contains the dependence on the spin variables in
eq.(\ref{average}). Droping this term, the sum over the spin configurations
yields simply $2^N$. As the remaining calculations are rather
straightforward we will only sketch them in the sequel.

To carry out the integrals over $v_i^k$ and $\hat{v}_i^k$ we
introduce the auxiliary parameters
$N q_k = \sum_i \left( \hat{v}_i^k \right)^2$,
$N m_k = \sum_i \hat{v}_i^k $, and $r_k = \sum_i v_i^k $.
After performing the resulting Gaussian integrals we introduce
the saddle-point parameters
$N M = \sum_i \hat{\Delta}_i$ and $N Q = \sum_i \hat{\Delta}_i^2$
which allow the decoupling of the indices $k$ and $i$.
The final result is
\begin{eqnarray}\label{n3}
\langle {\mathcal{N}} \left( \epsilon \right) \rangle &  =  & 2^N N^3 
\int \frac{dM d\hat{M}}{2 \pi} \int \frac{dQ d\hat{Q}}{2 \pi}
\int \frac{d\hat{\epsilon}}{2 \pi} \exp \left[ {\mathbf i} N \left( M \hat{M}
+ Q\hat{Q} + \epsilon \hat{\epsilon} \right) \right]
\nonumber \\
& & \times  \exp \left[ N \int_0^1 dz \ln  G_0 \left(z,\hat{\epsilon},M, Q \right)  + N
\ln G_1 \left(\hat{M},\hat{Q} \right)  \right]
\end{eqnarray} 
where
\begin{eqnarray}\label{G1}
G_0 & = & \int \frac{dq d\hat{q}}{2 \pi} \int \frac{dm d\hat{m}}{2 \pi} 
\int \frac{dr d\hat{r}}{2 \pi}
     \exp \left[ {\mathbf i} \, r \hat{r} -z \left( q + m^2 \right)
     -  {\mathbf i} \, \frac{\hat{\epsilon}}{8} r^2 \right]
\nonumber \\
& & \exp \left[ 
 {\mathbf i} \hat{m} \left(m - \hat{r} - r M \right) 
   + {\mathbf i} \hat{q} \left(q -  \hat{r}^2 - r^2 Q -
   2 \hat{r} r M  + 4{\mathbf i} M \right)
 \right] 
\end{eqnarray} 
and
\begin{equation}\label{G0}
G_1 =
\int  \frac{d \Delta d \hat{\Delta}}{2 \pi} \, \Theta \left( -\Delta \right)
\exp \left[ -  {\mathbf i} \hat{Q} \hat{\Delta}^2 +
 {\mathbf i} \hat{\Delta} \left( \Delta - \hat{M}\right)  
\right] .
\end{equation}
The integrals in eq.(\ref{n3}) are then evaluated in the limit $N
\rightarrow \infty$ by the standard saddle-point method, while the
integrals in the equations for $G_0$ and $G_1$ are trivially performed.
The final result for the exponent
\begin{equation}
\alpha \left( \epsilon \right) =
\frac{1}{N} \ln \langle {\mathcal{N}} \left( \epsilon \right) \rangle
\end{equation}
is simply
\begin{eqnarray}\label{alpha1}
\alpha \left( \epsilon \right) & = & {\mathbf i} \left[ 
\left( 2 +\hat{M} \right)  M  + Q\hat{Q} + 
\epsilon \hat{\epsilon} \right] + \ln \mbox{erfc} 
\left[ \frac{\hat{M}}{\left( 4 {\mathbf i} \hat{Q} \right)^{1/2} } \right]
\nonumber \\
& & -\frac{1}{2} \int_0^1 dz \ln \left[ 1 + 8 \left( Q - M^2 \right) z^2 
+ 8{\mathbf i} M z + {\mathbf i} \hat{\epsilon} z \right]
\end{eqnarray}
where the saddle-point parameters $M$, $\hat{M}$, $Q$, $\hat{Q}$, and
$\hat{\epsilon}$ are determined so as to maximize $\alpha$. In particular,
a brief analysis of the saddle-point equations indicates that $M$,
$\hat{Q}$ and $\hat{\epsilon}$ are imaginary so that $\alpha$ is real, as
expected. Introducing the real parameters $\mu = {\mathbf i} M$, $\beta =
{\mathbf i} \hat{\epsilon}$, $\eta = \hat{M}/ \left( 4 {\mathbf i} \hat{Q}
\right)^{1/2}$, and $\xi = -Q/M^ 2$, we rewrite eq.(\ref{alpha1}) as
\begin{eqnarray}
\label{alpha2}
\alpha \left(\epsilon \right) &=&
  2 \mu - \frac{\eta^2}{\xi} + \beta \epsilon   
  + \ln \mbox{erfc} \left( \eta \right) \nonumber \\
  & & -\frac{1}{2} \int_0^1 dz \ln \left[ 1 + \left( \beta + 8 \mu \right)z  +
8 \mu^2 \left( 1 + \xi \right) z^2  \right]
\end{eqnarray}
where we have used the saddle-point equation $\partial \alpha/ \partial \hat{Q} = 0$ 
to eliminate $\hat{Q}$. We note that in  eq.\ (\ref{alpha2})
the parameters $\eta$ and $\mu$ are decoupled which  facilitates greatly the
numerical problem of maximizing $\alpha$.

The number of local minima, regardless of their particular energy values,
is obtained by maximizing $\alpha$ with respect to $\epsilon$, which
corresponds to setting $\beta = 0$ in the saddle-point equations. In this
case, the value of the energy density that maximizes $\alpha$, denoted by
$\epsilon_t$, can be interpreted as the typical (average) energy density of
the local minima.  We find $\alpha=0.4394$ and $\epsilon_t=0.0837$. These
results agree very well with the numerical data $\alpha\approx0.4388\pm(7)$
and $\epsilon_t \approx0.0826\pm(6)$, obtained through the exhaustive search for
$N\leq 20$ and averaging over $100$ realizations of the
couplings.

Moreover, an exhaustive search for $N\leq 30$ yields that the exponent
governing the exponential growth of the number of local minima in the pure
model ${\mathcal{H}}$ is $0.453\pm(7)$ and the typical energy density of
the minima is $0.086 \pm(2)$. Hence, so far as the statistics of metastable
states is concerned, the mean-field Hamiltonian ${\mathcal{H}}_d$ yields in
fact a very close approximation to the pure Hamiltonian ${\mathcal{H}}$.
For the purpose of comparison we note that 
$\alpha = 0.1992$ and $\alpha = 0.3552$ for the 
binary $\pm 1$ 2-spin glass \cite{Tanaka:80a,Bray:81} and 4-spin glass 
models \cite{Gross:84,Stadler:96a}, respectively, while
$\alpha = \ln 2\approx 0.6931$ for the random energy model
\cite{Derrida:80a}.

\begin{figure}[t]
\par
\centerline{\epsfig{file=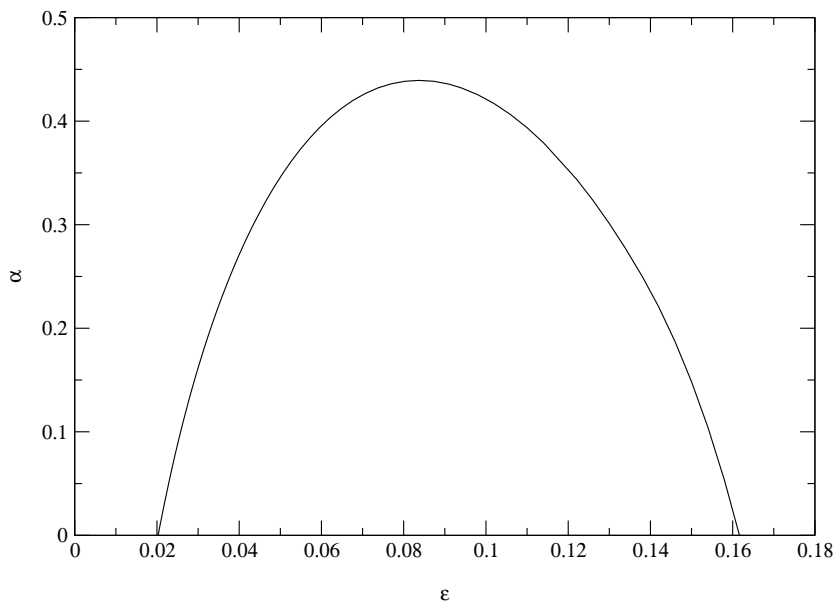,width=0.70\textwidth,clip=}}
\par
\caption{Exponent $\alpha(\epsilon)$ as a function of the
energy density $\epsilon$. The ground state energy is defined by
$\alpha(\epsilon_0)=0$. We find $\epsilon_0=0.0202845$.}
\label{fig:Ne}
\end{figure}

In Fig.\ref{fig:Ne} we show the exponent $\alpha$ as a function of the
energy density $\epsilon$. For the sake of clarity we present only the
region of positive values of $\alpha$.  The lowest value of $\epsilon$ at
which the exponent $\alpha$ vanishes, denoted by $\epsilon_0$, gives a
lower bound to the ground-state energy density of the spin model defined by
the Hamiltonian (\ref{eq:H_BM}) \cite{Tanaka:80a}.  We find $\epsilon_0 =
0.0202845$ which, within the numerical precision, is exactly the value
predicted by the first step of replica-symmetry breaking \cite{Bouchaud:94}
as well as by Golay's ergodicity hypothesis \cite{Golay:77,Bernasconi:87}.
This coincidence between the
replica and the density of metastable states predictions for the
ground-state energy occurs also in the random energy model
\cite{Derrida:80a,Gross:84}. 

A similar study of the symmetrized
version of the mean-field Hamiltonian (\ref{eq:H_BM}), in which $J_{ij}^k =
J_{ji}^k$, yields exactly the same expression for the exponent $\alpha$,
see eq.(\ref{alpha2}), provided that the energy density $\epsilon$ is
replaced by $\epsilon_s/2$.  Hence the symmetrization procedure results in
a trivial rescaling the energy densities of the local minima, without
affecting their number.

\section{Energy Barriers and Basin Sizes}
\label{sect:barriers}

The picture that comes out of the replica approach to disordered
spin models is that the phase space $V$ composed of the
$2^N$ spin configurations is broken into several valleys 
 connected by saddle points \cite{Mezard:87}.
The relative location and energetic properties of valleys and saddles
are expected to determine e.g.\ the ease with which the ground state
can be reached.

It will be convenient to introduce the notion of saddle-point energy
$E[s,w]$ between two (not necessarily strict) minima $s$ and $w$. Denoting,
for the sake of generality, the energy of state $x$ by $f(x)$, we can write
\begin{equation}
  E[s,w] = \min\left\{ \max\left[ f(z) \big| z\in\mathbf{p} \right]
             \,\bigg|\,
            \mathbf{p}: \textrm{path from } s \textrm{ to }  w
                            \right\} ,
\label{eq:saddle}
\end{equation}
where a path $\mathbf{p}$ is a sequence of configurations connected
by one-spin flips (or, more generally, by moves taken from any desired
``move set''). The 
saddle-point energy $E[s,w]$ forms an ultrametric distance measure
on the set of local minima, see e.g.\ \cite{Rammal:86,Vertechi:89}.
The {\em barrier} enclosing a local minimum is the height of the lowest
saddle point that gives access to an energetically more favorable 
minimum. In symbols:
\begin{equation}
  B(s) = \min\left\{ E[s,w] - f(s) \big| w:f(w)<f(s) \right\} 
\label{eq:defdepthC}
\end{equation}
If $B(s)=0$ then the local minimum  $s$ is marginally stable. It is easy to
check that eq.(\ref{eq:defdepthC}) is equivalent to the definition of the
depth of local minimum in \cite{Kern:93}. It agrees for metastable states
with the more general definition of the depth of a ``cycle'' in the
literature on inhomogeneous Markov chains
\cite{Azencott:92,Catoni:92,Catoni:99}.

The information contained in the energy barriers is conveniently summarized
by two global parameters that e.g.\ determine the convergence behavior of
Simulated Annealing and related algorithms. 
The {\em depth} of a landscape \cite{Hajek:88,Catoni:92,Kern:93,Ryan:95}
is defined as
\begin{equation}
  \mathsf{D} = \max\left\{B(s) \big|
       s \textrm{ is not a global minimum } \right\} .
\label{eq:Depth}
\end{equation}
It can be shown that Simulated Annealing converges almost surely to a
ground state if and only if the cooling schedule $T_k$ satisfies
$\sum_{k\ge0}\exp(-\mathsf{D}/T_k)=\infty$ \cite{Hajek:88}. In order to
make the depth comparable between different landscapes we shall consider
below the dimensionless parameter $\mathsf{D}/\sigma$, where $\sigma^2$ is
the variance of the energy across the landscape.
A related quantity is the (dimensionless) {\em difficulty}
\cite{Catoni:92,Catoni:99} of the
landscape, defined by 
\begin{equation}
  \psi = \max\bigg\{ \frac{B(s)}{f(s)-f(\min)} \bigg|
         s\textrm{ is not a global minimum}\bigg\} 
\label{eq:difficulty}
\end{equation}
where $f(\min)$ is the global energy minimum and the maximum is taken over
non-global minima only. It is directly related to the optimal
speed of convergence of Simulated Annealing.

Since a direct evaluation of eq.(\ref{eq:saddle}) would require the
explicit constructions of all possible paths it does not provide a feasible
algorithm for determining $E[s,w]$ even if $N$ is small enough to allow an
exhaustive survey of the landscape. The values of $E[s,w]$ and $B(s)$ can,
however, be retrieved from the {\em barrier tree} of the landscape.
Barrier trees have been considered recently in the context of RNA folding
\cite{Flamm:00a} and under the name ``disconnectivity graphs'' in the
protein folding literature \cite{Becker:97,Garstecki:99}. In this
contribution we use a modified version of the program {\tt barriers},
which was developed for the analysis of RNA folding landscapes in
\cite{Flamm:00a}. For the sake of completeness we briefly outline the
definition of the barrier trees below.

\begin{figure}[t]
\par
\centerline{\epsfig{file=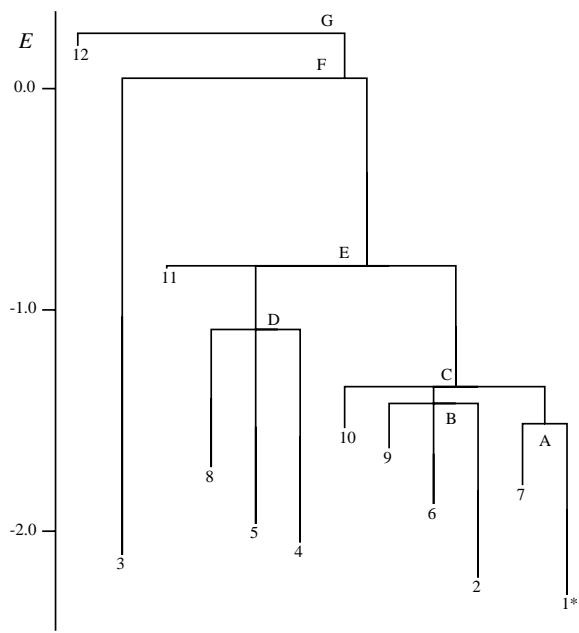,width=0.5\textwidth}}
\par
\caption{Example of a barrier tree of a landscape. Data belong to a
Gaussian {\tt REM} with $N=7$. The leaves 1-12 denote the local minima. The
global minimum 1 is marked with an asterisk. Saddle points are labelled
with capital letters from A to G. The saddle points $\mathsf{B}$,
$\mathsf{C}$, $\mathsf{D}$, $\mathsf{E}$ are ``degenerate'' indicating that
the lowest energy paths leaving e.g.\ 4,5,8 run through a common exit
point. (Note that all $2^7=128$ configurations have pairwise distinct
energies, hence there are no two distinct saddle points with the same
energy, which may exist e.g.\ in the {\tt LABSP}.) The Barrier of $5$ is
$B(5)=E(\mathsf{D})-E(5)$, along the lowest path from $5$ to $4$, while
$B(4)=E(\mathsf{E})-E(4)$, along the lowest path from $4$ to $1^*$.}
\label{fig:STree}
\end{figure}

For simplicity let us assume that the energies of any two spin
configurations are distinct, i.e., there is a unique ordering of the
spin configurations by their energies. The construction of the barrier
tree starts from an energy-sorted list of all configurations in the
landscape. We will need two lists of valleys throughout the calculation:
The global minimum $x[1]$ belongs to the first active valley $V_1$,
while the list of inactive valleys is empty initially.  
Going through this list of all configurations in the order of increasing
energy we have three possibilities for the spin configuration $x[k]$
at step $k$.
\begin{itemize}
\item[(i)] $x[k]$ has neighbors in exactly one of the active valleys $V_i$.
   Then $x[k]$ belong to $V_i$.
\item[(ii)] $x[k]$ has no neighbor in any of the (active or inactive)
   valleys that we have found so far. Then $x[k]$ is a
   local minimum and determines a new active valley $V_{l}$.
   In the barrier tree $x[k]$ becomes a leaf.
\item[(iii)] $x[k]$ has a neighbor in more than one active valley, say
   $\{V_{i_1},V_{i_2},\dots,V_{i_q} \}$. Then it is a saddle point
   connecting these active valleys. In the barrier tree $x[k]$ becomes
   an internal node. In this case we add $x[k]$ to
   valley $V_{i_1}$ with the lowest energy. Then we copy the configurations
   of $V_{i_2},\dots,V_{i_q}$ to $V_{i_1}$. Finally, the status of 
   $V_{i_2}$ through $V_{i_q}$ is changed from active to inactive.
   This reflects the fact that from the point of view of a configuration
   with an energy higher than the saddle point $x[k]$,
   $V_{i_1},\dots,V_{i_q}$ appear as a single valley that is subdivided
   only at lower energy. Consequently, after the highest saddle-point
   energy has been encountered, all valleys except for the globally
   optimal $V_1$ are in the inactive list.
\end{itemize}
The outcome of this procedure is a tree such as the one shown in
Fig.~\ref{fig:STree}. The leaves correspond to the valleys of the
landscape, while the interior nodes denote the saddle points. The tree
contains the information on all local minima and their connecting saddle
points.  Indeed, saddle-point energies, and energy barriers can be
immediately read off the barrier trees.

\begin{figure}[t]
\par
\centerline{\epsfig{file=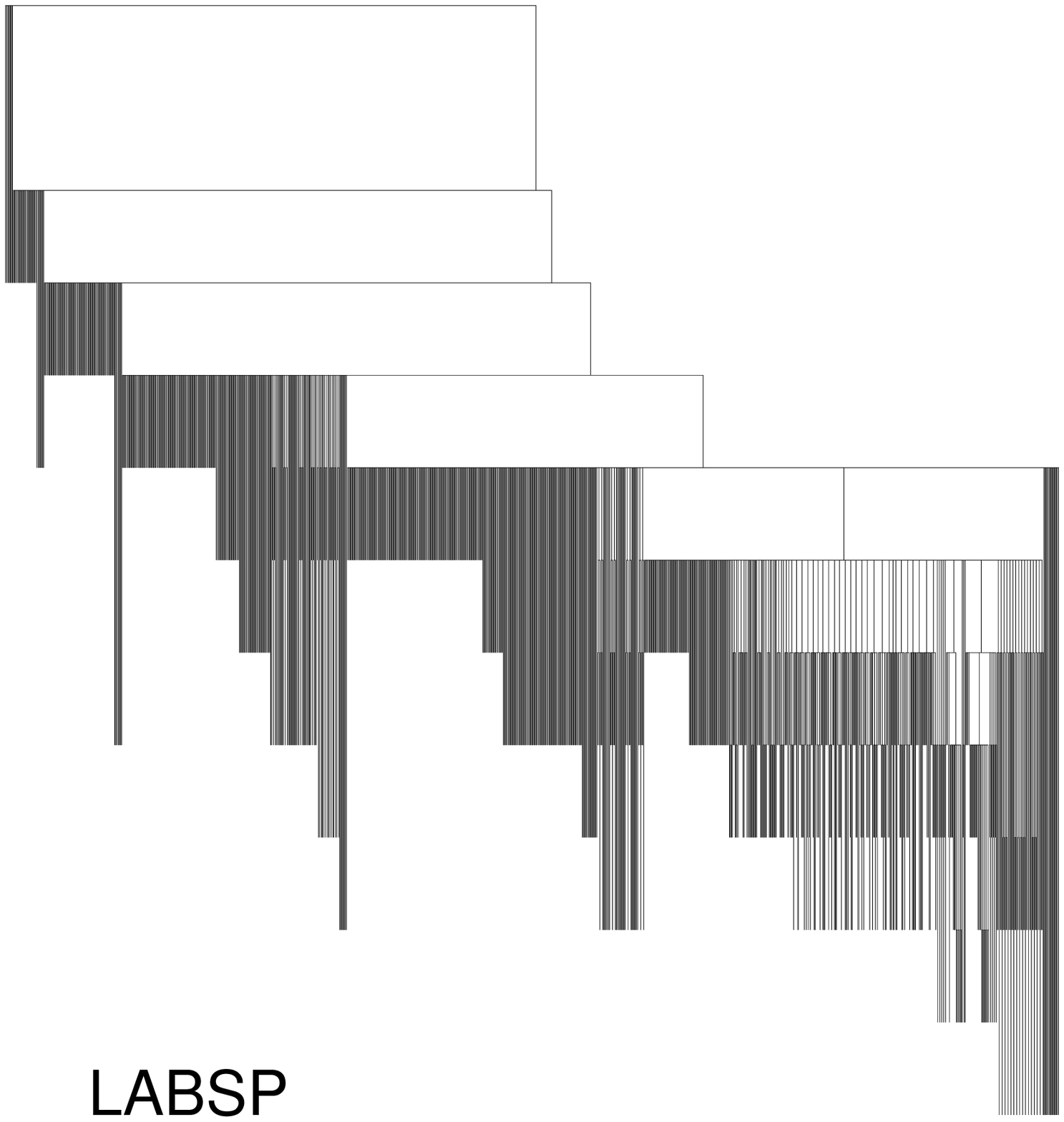,width=0.4\textwidth}
\qquad\qquad
\epsfig{file=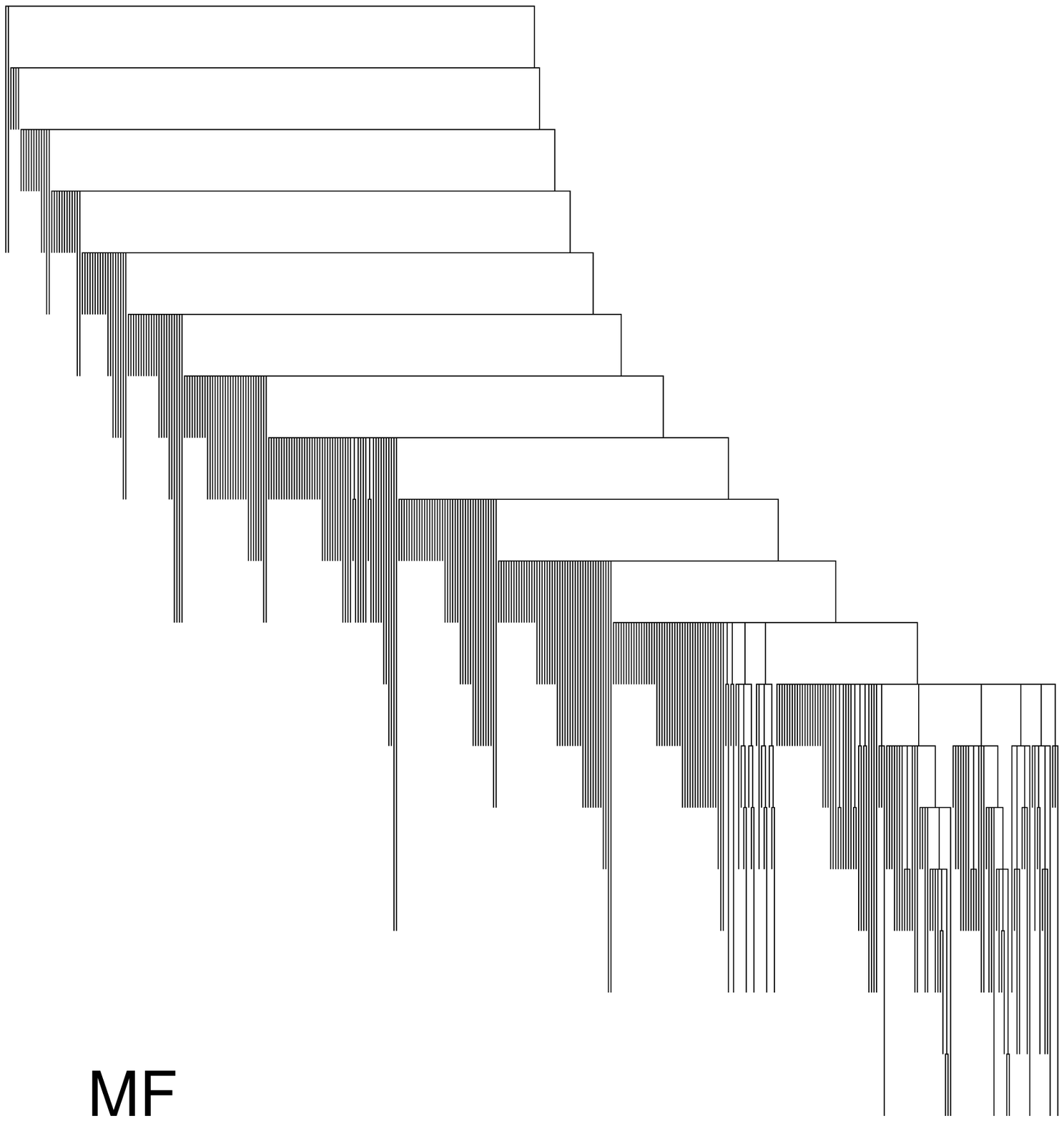,width=0.4\textwidth}}
\par
\centerline{\epsfig{file=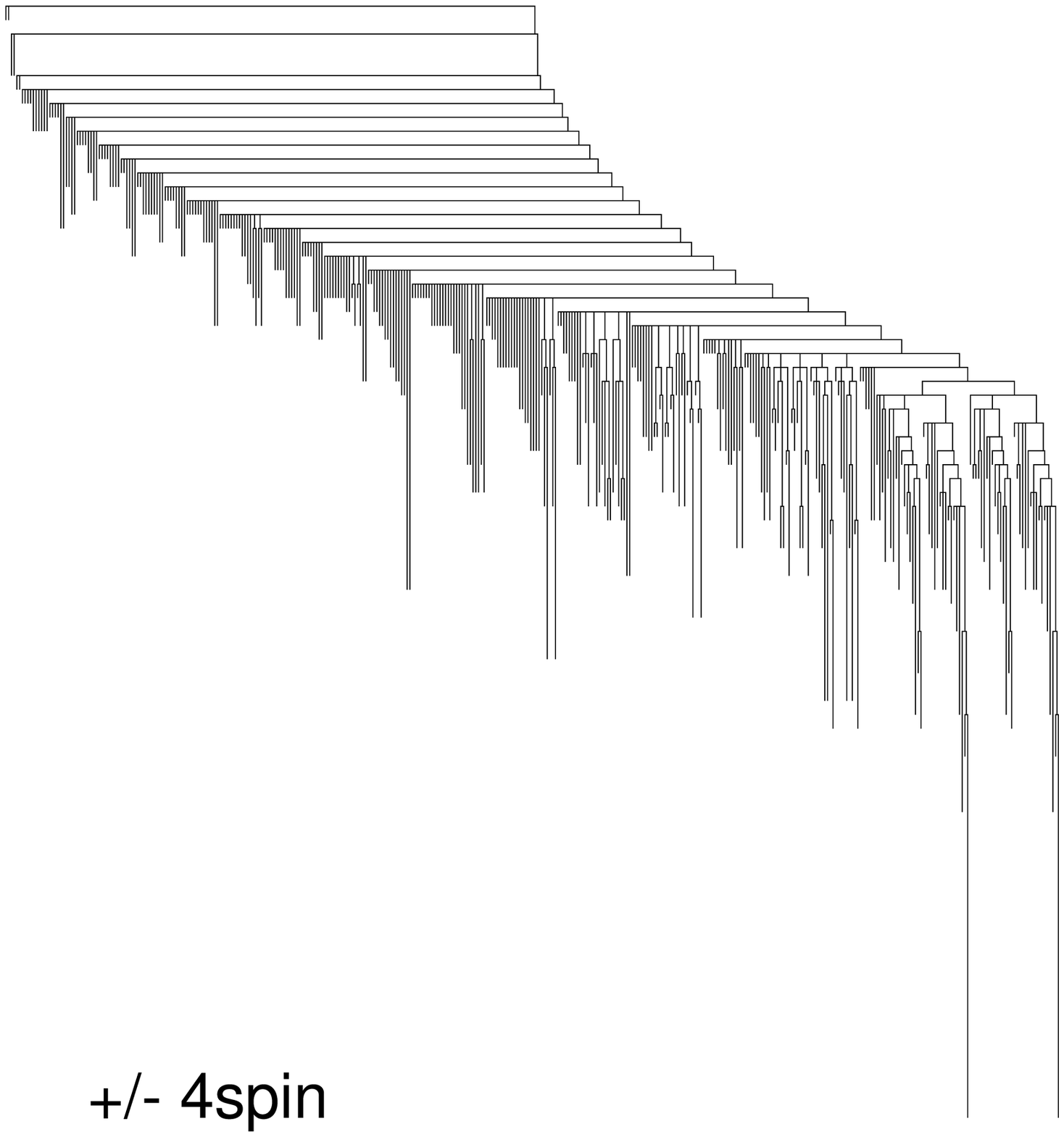,width=0.4\textwidth}
\qquad\qquad
\epsfig{file=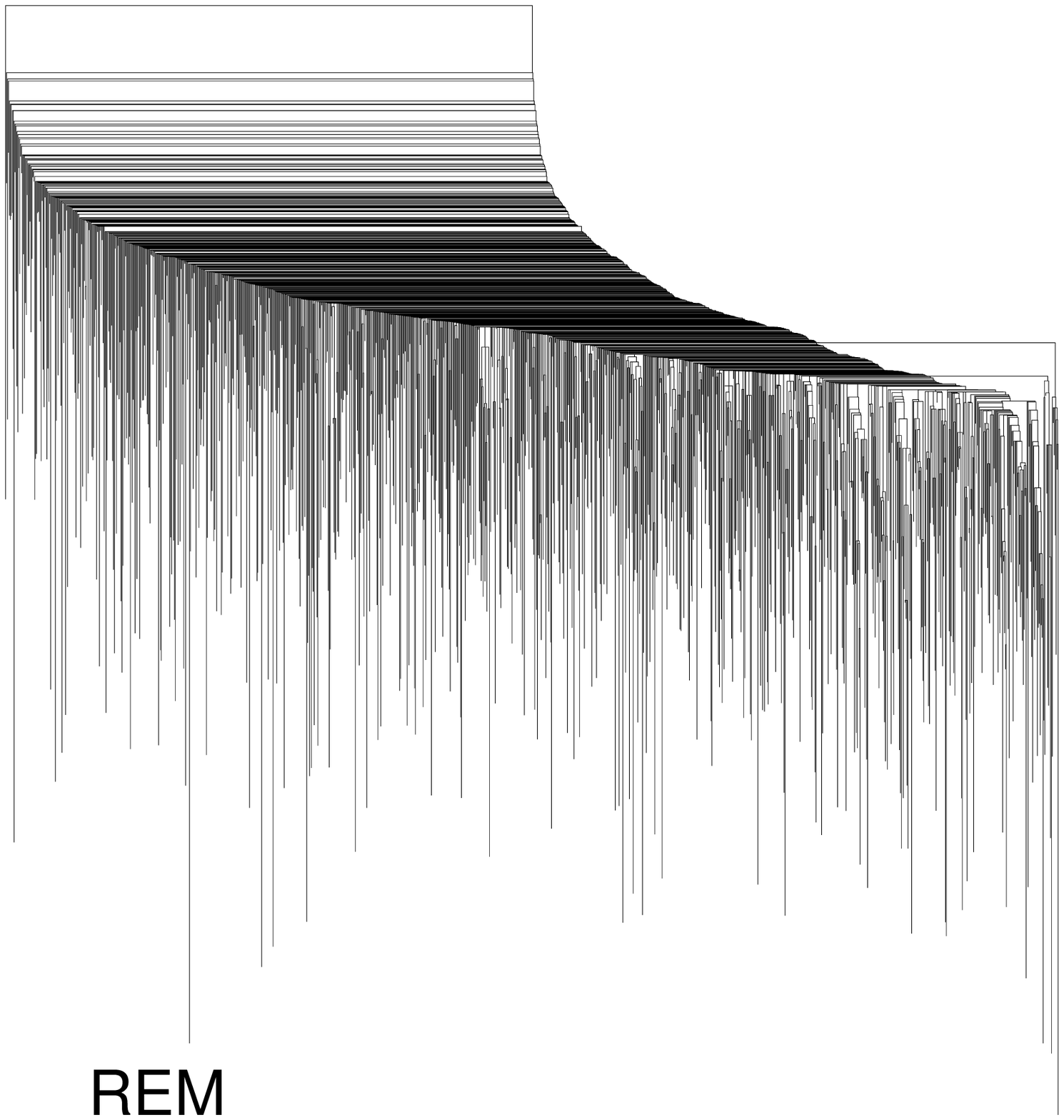,width=0.4\textwidth}}
\par
\caption{Tree representation of typical landscapes with $N=16$. Upper left
{\tt LABSP}, upper right mean-field approximation, lower left 
integer 4-spin model, lower right Gaussian {\tt REM} for $N=14$ since for $N=16$ the
number of minima is too large to allow a meaningful drawing.}
\label{fig:trees}
\end{figure}

A precise definition of {\em valleys} and {\em saddle points} in a
landscape requires that we take into account the degeneracies in the energy
function, i.e., the existence of distinct spin configurations with
identical energies, in particular, the presence of neutrality, where
neighboring configurations have identical energies \cite{Reidys:00a}.
Degeneracies complicate the construction of the barrier tree, since the
energy-sorting of the landscape is not unique any more. The simplest remedy
is to use the same procedure as above starting from an arbitrary energy
sorting. In this case the order of degenerate configurations in the list is
arbitrary but fixed throughout the computation. Before proceeding to a
configuration with strictly higher energy a simple clean-up step needs to
be included in the tree-building algorithm: adjacent valleys with
$E[s,w]=f(s)=f(w)$ are joined to a single valley. Note that the resulting
barrier tree may still contain distinct valleys with the same energy, as
the examples in Fig.~\ref{fig:trees} show. The leaves of the barrier trees
are in general valleys which may contain more than one degenerate local
minimum.

There is a clear visual difference between the barrier trees for {\tt
LABSP} and the mean-field approximation {\tt MF} at the one hand, and the
$\pm1$ 4-spin Hamiltonian and the {\tt REM} on the other hand. The main
difference appears to be a much larger amount of degeneracy in {\tt
LABSP}/{\tt MF}, in particular highly degenerate ground states.  In fact,
it can be shown that the pure Hamiltoninan (\ref{eq:H}) has many nontrivial
symmetries, besides the trivial one where $x$ is replaced by $-x$, which
are then responsible for the high degeneracy observed in the tree barrier
\cite{Mertens:96}. Obviously, the disordered Hamiltonian (\ref{eq:H_BM})
cannot have the same symmetries as the pure one, and so its high degeneracy
stems simply from the extreme dilution of the couplings $J_{ij}^k$.  All
models, except {\tt REM}, are symmetric under replacing $x$ by $-x$, hence
all states appear in pairs.  We note that the barrier tree of the $\pm1$
4-spin model is reminiscent of the ``funnels'' discussed e.g.\ in
protein folding, with a large energy difference between the two global
optima and almost all local ``traps''. In contrast, the {\tt REM} shows, as
expected, no relationship between energy and nearness of local minimum to
the global one.

\begin{figure}[t]
\par
\centerline{\psfig{file=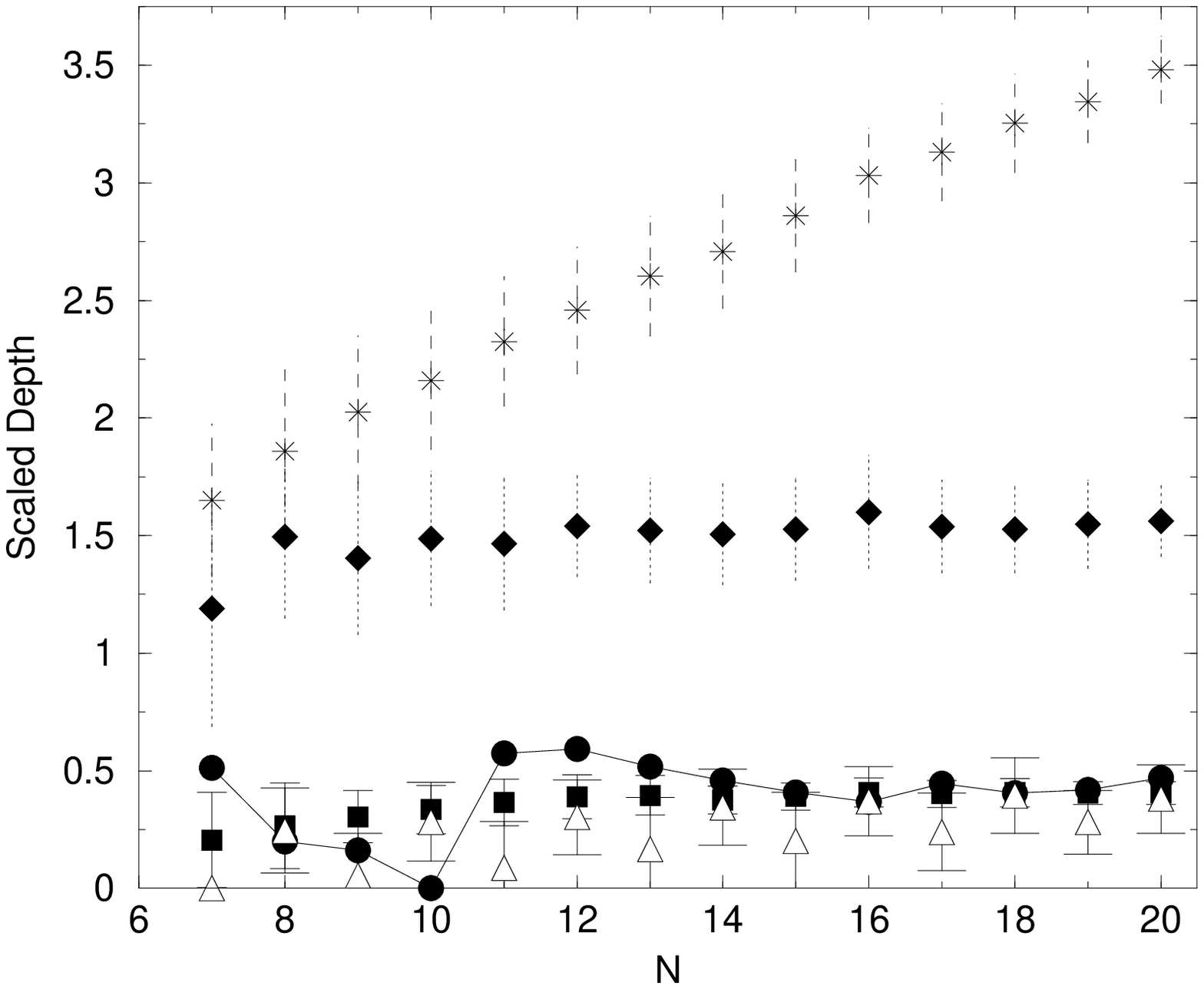,width=0.47\textwidth}\hfill
            \psfig{file=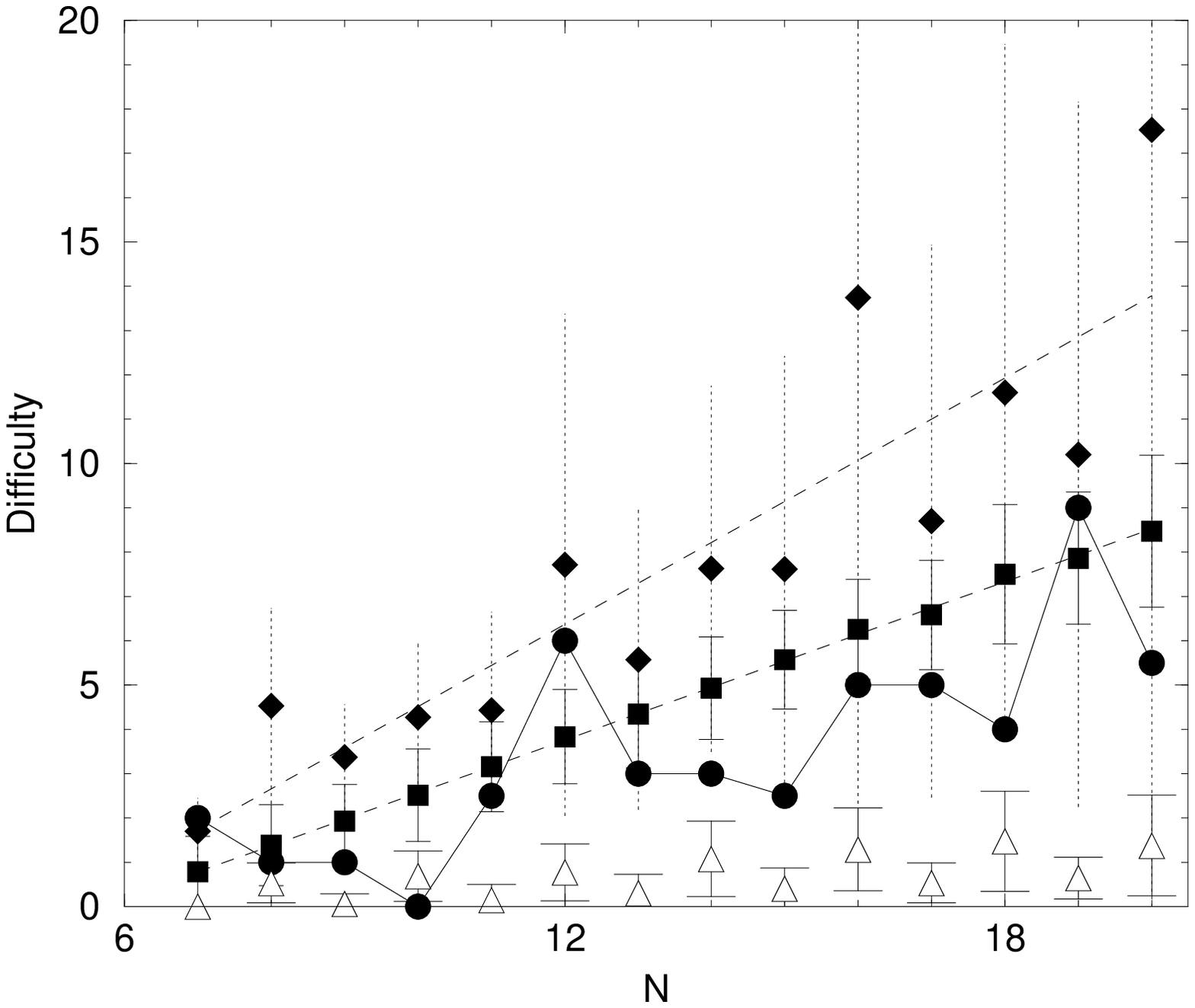,width=0.47\textwidth}}
\par
\caption{Depth and Difficulty. Symbols:
$\bullet$ {\tt LABSP},
$\blacksquare$ mean-field approximation,
$\triangle$ $\pm1$-version of the $2$-spin model,
$\blacklozenge$ $\pm1$-version of the $4$-spin model,
$\star$ Gaussian {\tt REM}. Data are averaged over 100 instances (50 instances
for $N=20$); the error bars show the width of the distribution, not the
standard error of the means.\hb
}
\label{fig:DD}
\end{figure}

During the construction of the barrier tree it is easy only to compute the
lowest barrier $B'(s)$ from $s$ to a local minimum that comes earlier in
the list of configurations, instead of the lowest barrier $B(s)$ to a local
minimum with strictly smaller energy. Clearly, $B'(s)\le B(s)$ since we
take the minimum over a few more configurations than prescribed by
eq.(\ref{eq:defdepthC}). In case of degenerate landscapes our version of
the {\tt barriers} program calculates $B'(s)$ which depends on the ordering
of the degenerate configurations.  We obtain, however, $B(s)=B'(s)$ for at
least one of the valleys at each energy level. The fact that in
eq.(\ref{eq:Depth}) we are required to maximize in particular over the
barriers necessary to escape from any given energy level, however, implies
that the values of depth and difficulty can be obtained directly from
$B'(s)$ instead of $B(s)$. We note at this point that a modified version
$\mathsf{D}'\ge\mathsf{D} $ of the depth in which the maximum over all
non-global minima is replaced by the maximum over all minima except one
global minimum $x^*$ can also be obtained by the simplified procedure
above, since it can be shown that $\mathsf{D}'$ is independent of the
choice of $x^*$ \cite{Catoni:99}. The parameter $\mathsf{D}'$ also appears
in exact results on the convergence of Simulated Annealing.

Depth and Difficulty are shown in Figure~\ref{fig:DD} as a function of the
number $N$ of spins.  While there are (moderate) quantitative differences,
there does not seem to be a qualitative difference between the {\tt LABSP},
the mean-field approximation, and the discretized 2 and 4-spin models.  Note that
all landscapes with the exception of the Gaussian {\tt REM} have constant
scaled depth $\mathsf{D}/\sigma$, while $\mathsf{D}/\sigma$ increases
linearly with the system size in the {\tt REM}.

A linear regression of the difficulties yields the slopes $0.595\pm6$,
$0.926\pm14$, and $0.07\pm2$ for the mean-field Hamiltonian, the
$\pm1$-version of the $4$-spin model and the $\pm 1$ 2-spin model,
respectively.  As expected, the difficulty of the quadratic model is much
smaller than the difficulty of the 4-spin model.  For the sake of clarity,
we have omitted the data about the mean difficulty of the {\tt REM} since
it is too large as compared to those shown in the figure. Moreover the
width of its difficulty distribution is also so large that the mean value
is not physically meaningful.

\begin{figure}[t]
\par
\centerline{\psfig{file=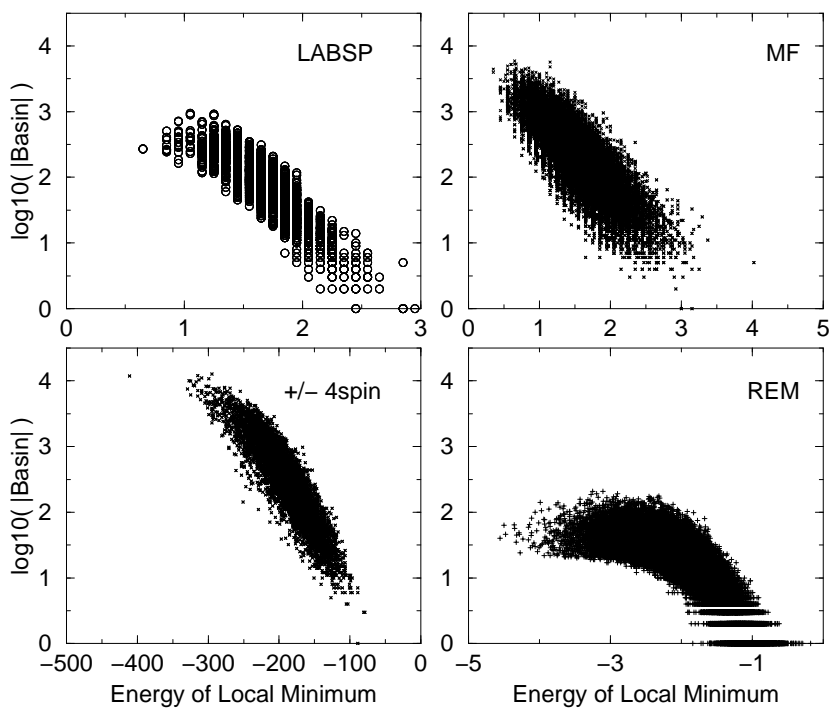,width=0.70\textwidth}}
\par
\caption{The logarithm of the size of the gradient-walk basin
of attraction of the minima as function of their energies for $N=20$. The
data for {\tt MF} and the $\pm1$ 4-spin model are superpositions of 10
instances, while for the {\tt REM} we show only a single instance.
}
\label{fig:Basins}
\end{figure}

Additional information on the local minima can be traced during the
construction of the barrier tree. We say that a configuration $x$ belongs
to the basin of the local minimum $s$ if $s$ is the endpoint of the {\em
gradient walk} (steepest descent) starting in $x$. (Recall that each step
of a gradient walk goes to the lowest energy neighbor.)  By determining the
valley to which the lowest energy neighbor of $x[k]$ belongs we may for
instance record the {\em basin size} of each local minimum. In a landscape
without neutrality the gradient walk is uniquely determined by the initial
condition, hence the basins form a partition of the set of
configurations. We neglect the effects of neutrality in our numerical data
by directing the gradient walk to the first possibility in the
energy-sorted list of configurations.

Computationally we find, for all models but the {\tt REM}, that there is an
approximate linear relationship between the energy of a local minimum and
the logarithm of the size of its gradient-walk basin of attraction, see
Figure~\ref{fig:Basins}. The fact that the deepest valleys have the largest
basins of attraction can be understood as a consequence of the correlation
between neighboring spin configurations in all landscapes with the
exception of the {\tt REM}, for which all low-energy minima have
essentially the same size of basin of attraction.

\section{Discussion}
\label{sect:disc}

The performance evaluation of local search heuristics, in particular
Simulated Annealing, in typical instances of optimization problems is a
relatively new subject, where the existing criteria for measuring the
hardness or difficulty of a problem are still not widely known or accepted,
as compared to e.g.\ the more traditional worst-case analysis.  In fact, on
the one hand, one expects that the average number of local minima may serve
as a measure of the problem hardness, while, on the other hand, one must
concede that only local minima separated by high energy barriers are
potential traps for the search heuristic. In this paper we combine the
concepts of depth and difficulty from the Simulated Annealing literature to
the average density of states calculations from the statistical mechanics
of disordered systems to obtain a reasonably complete statistical
description of the energy landscapes of several classic {\em disordered}
models. The motivation is to compare the statistical features of these
landscapes with the properties of a rather puzzling {\em deterministic}
problem --- the low autocorrelated binary string problem ({\tt LABSP}) ---
which has been identified as a particularly hard optimization problem for
search heuristics such as Simulated Annealing.

Our results indicate that there is only a quantitative difference between
the depths and difficulties, as defined in the Simulated Annealing
literature, of all models investigated, with the exception of the random
energy model ({\tt REM}) for which the complete lack of correlations
between the energies of neighboring configurations results in a genuine
golf-course type landscape. Hence, we have found no evidences of a
golf-course like structure in the {\tt LABSP} landscape, which resembles
much more a correlated spin-glass model than the {\tt REM}.  It must be
emphasized that although the pure {\tt LABSP} model (\ref{eq:H}) may have a
glass phase characterized by uncorrelated equilibrium states (at least its
mean-field version has such a phase \cite{Bouchaud:94}), the mere existence
of this phase is no evidence of a golf-course like structure which, as
mentioned above, requires vanishing correlations between the energy values
of neighboring spin configurations.

Perhaps the ``golf-course'' conjecture \cite{Bernasconi:87,Mertens:96}
stems simply from the fact that for large $N$ the {\tt LABSP} is a much
more difficult problem for Simulated Annealing than the familiar quadratic
spin glass, as shown in Fig.\ \ref{fig:DD}.  Interestingly, the pairwise
comparison between the problems indicates that those problems with the
larger number of local minima have also the larger difficulty, the only
exception being {\tt LABSP} and the 4-spin glass model. It should therefore
be interesting to use these two problems as a test-bed for validating the
hardness criteria proposed in the Simulated Annealing literature.

Finally, our analysis has shown that the disordered, mean-field Hamiltonian
${\mathcal{H}}_d$, eq.(\ref{eq:H_BM}), describes surprisingly well the
qualitative (e.g.\ the barrier trees) as well as the quantitative (e.g.\
number and typical energy of local minima) features of the pure model
${\mathcal{H}}$, eq.(\ref{eq:H}).

\subsection*{Acknowledgements}

Special thanks to Christoph Flamm for the source code of his program {\tt
barriers}.  The work of JFF was supported in part by Conselho Nacional de
Desenvolvimento Cient{\'\i}fico e Tecnol{\'o}gico (CNPq). The work of PFS
was supported in part by the Austrian Fonds zur F{\"o}rderung der
Wissenschaftlichen Forschung, Proj.\ No.\ 13093-GEN. FFF is supported by
Funda\c{c}\~ao de Amparo \`a Pesquisa do Estado de S\~ao Paulo (FAPESP). We
thank FAPESP for supporting PFS's visit to S\~ao Carlos, where part of his
work was done.

\section*{References}

\bibliographystyle{abbrv}
\bibliography{labsp}


\end{document}